\documentclass{emulateapj}
\usepackage{apjfonts}
\usepackage{graphicx}
\usepackage[dvips]{color}

\shorttitle{ERE Excitation}
\shortauthors{Witt et al.}

\begin{document}

\title{The Excitation of Extended Red Emission: New Constraints on its Carrier from HST Observations of NGC 7023\altaffilmark{1}}

\author{Adolf N.\ Witt\altaffilmark{2},
        Karl D.\ Gordon\altaffilmark{3},
        Uma P.\ Vijh\altaffilmark{2},
        Paul H.\ Sell\altaffilmark{2},
        Tracy L.\ Smith\altaffilmark{4}, \&
        Rui-Hua Xie\altaffilmark{5}}

\altaffiltext{1}{Based on observations made with the NASA/ESA Hubble Space Telescope, obtained at the Space Telescope Science Institute, which is operated by the Association of Universities for Research in Astronomy, Inc., under NASA contract NAS 5-26555. These observations are associated with program \#9471.}
\altaffiltext{2}{Department of Physics \& Astronomy, The University of Toledo, Toledo, OH 43606, (awitt@dusty.astro.utoledo.edu, uvijh@astro.utoledo.edu, psell@utnet.utoledo.edu)}
\altaffiltext{3}{Steward Observatory, University of Arizona, Tucson, AZ 85721 (kgordon@as.arizona.edu)}
\altaffiltext{4}{Space Science Institute Columbus, Smith Laboratory, Room 2192, Department of Physics, Ohio State University, Columbus, OH 43210 (tsmith@campbell.mps.ohio-state.edu)}
\altaffiltext{5}{Department of Physics \& Institute for Quantum Studies, Texas A\&M University, 4242 TAMU, College Station, TX 77843 (rhxie04@tamu.edu)}

\begin{abstract}
The carrier of the dust-associated photoluminescence process causing the extended red emission (ERE) in many dusty interstellar environments remains unidentified. Several competing models are more or less able to match the observed broad, unstructured ERE band. We now constrain the character of the ERE carrier further by determining the wavelengths of the radiation that initiates the ERE. Using the imaging capabilities of the Hubble Space Telescope, we have resolved the width of narrow ERE filaments appearing on the surfaces of externally illuminated molecular clouds in the bright reflection nebula NGC 7023 and compared them with the depth of penetration of radiation of known wavelengths into the same cloud surfaces. We identify photons with wavelengths shortward of 118~nm as the source of ERE initiation, not to be confused with ERE excitation, however. There are strong indications from the well-studied ERE in the Red Rectangle nebula and in the high-$|b|$ Galactic cirrus that the photon flux with wavelengths shortward of 118 nm is too small to actually excite the observed ERE, even with 100\% quantum efficiency. We conclude, therefore, that ERE excitation results from a two-step process. The first step, involving far-UV photons with E~$>$~10.5~eV, leads to the creation of the ERE carrier, most likely through photo-ionization or photo-dissociation of an existing precursor. The second step, involving more abundant near-UV/optical photons, consists of the optical pumping of the previously created carrier, followed by subsequent de-excitation via photoluminescence. The latter process can occur many times for a single particle, depending upon the lifetime of the ERE carrier in its active state.  While none of the previously proposed ERE models can match these new constraints, we note that under interstellar conditions most polycyclic aromatic hydrocarbon (PAH) molecules are ionized to the di-cation stage by photons with E~$>$~10.5~eV and that the electronic energy level structure of PAH di-cations is consistent with fluorescence in the wavelength band of the ERE. Therefore, PAH di-cations deserve further study as potential carriers of the ERE.
\end{abstract}


\section{Introduction}
\subsection{Definitions}

Extended Red Emission (ERE) results from a dust-related optical photoluminescence process in the interstellar medium. Such processes  are excited through the absorption of  higher-energy photons by a suitable carrier, followed by emission of photons at lower energy. If the photoluminescence arises  from such a single-step process, we will refer to the energizing process as $\emph{excitation.}$ In some instances, the carrier itself is the result of an earlier photo-process,  e.g. photo-ionization or photo-dissociation, which by itself does not result in luminescence. Since the creation of this carrier is a necessary precondition for later photoluminescence, we will call the enabling process $\emph{initiation.}$ In this paper, we will investigate whether the ERE is the result of a single-step photo-excitation or the consequence of a two-step process, in which ERE initiation is followed by ERE excitation.

\subsection{Background}

The ERE was first observed in the peculiar bi-polar Red Rectangle nebula \citep{Cohen75, Schmidt80}. Soon after its initial discovery, ERE was shown to be present in many other dusty interstellar environments, albeit at mostly lower intensities, e.g. in reflection nebulae \citep{Witt84, Witt90}, HII regions \citep{Perrin92, Darbon00}, carbon-rich planetary nebulae \citep{Furton90, Furton92}, as well as  the diffuse interstellar medium of the Milky Way \citep{Gordon98} and other galaxies \citep{Perrin95, Pierini02}.  The ERE intensity in many dusty sources is proportional to the local density of the illuminating radiation field. This provides a strong argument in favor of the suggestion that the ERE is a photoluminescence process. 

The spectroscopic signature of the ERE is a broad (60 to 100~nm FWHM), unstructured emission band, typically extending from 540~nm to beyond 900~nm in wavelength. The peak wavelength of the ERE band varies from somewhat longward of 600~nm to beyond 800~nm in response to varying environmental conditions, in particular the density of the illuminating ultraviolet (UV) radiation field \citep{Smith02}. This is a defining characteristic of the ERE which distinguishes it from other emission features with essentially invariable emission wavelengths.

 At high Galactic latitudes, the intensity of the ERE is comparable to that of the dust-scattered diffuse galactic light (DGL), which has led to an estimated lower
limit of the ERE quantum yield of $10 \pm 3$\% and the conclusion that the ERE carrier must be a major contributor to the absorption part of interstellar extinction at UV/visible wavelengths in the Milky Way \citep{Gordon98}. Its ubiquitous presence in radiation environments ranging in density over more than five orders of magnitude testifies to the relative robustness of the ERE carrier. More details about ERE studies done over the past three decades may be found in a recent review by \citet{Witt04}.

\subsection{Past Studies of ERE Excitation/Initiation}

None of the current models for interstellar dust \citep[e.g.][]{Draine04, Zubko04} predict ERE, nor do they provide a satisfactory post-facto explanation for the observed characteristics of the ERE. As reviewed by \citet{Witt04}, numerous ad-hoc models have been proposed to account for the ERE. Most of them suggest ERE carriers in the form of large molecular structures or nanometer-sized grains, in which electronic excitations by shorter-wavelength photons are followed by efficient electronic radiative transitions across a bandgap of $< 2$~eV. Well-studied examples of such processes are fluorescence and phosphorescence in organic molecules and the photoluminescence in semiconductor nanoparticles. Such particles may be able to meet the constraints posed by the observed spectral characteristics, including their environment-dependent variations, and by the inferred quantum efficiency \citep{Smith02}.  However, all carrier models proposed so far suffer from the following dilemma.

The excitation of fluorescence or photoluminescence generally occurs with optimum efficiency at energies just slightly above the bandgap energy.  The difference between the optimum excitation energy and the energy at which the luminescence emerges, the so-called Stokes shift, is typically of the order of 1~eV or less in typical photoluminescent agents. At first glance, this does not appear to be the case for the ERE in view of two previous studies of the excitation requirements of the ERE. The first study by \citet{Witt85}, based upon multi-band surface brightness photometry of several reflection nebulae, concluded that either near-UV photons in the vicinity of the 217.5~nm absorption peak in the extinction curve and/or far-UV photons at wavelengths shortward of 150~nm provided  the correct match for the required excitation to result in the observed spatial distribution of the ERE in these objects. These results provided the first indication  that ultraviolet rather than optical photons are responsible for ERE initiation/excitation.

The second study related to ERE excitation by \citet{Darbon99} compiled the records of positive ERE detections as well as the results of negative searches for ERE in a large number of nebulae and related them to the effective temperatures of the exciting stars. While ERE is present in abundance (with few exceptions) among nebulae illuminated by stars with $T_\mathrm{eff} > 10,000~K$, not a single nebula with stars of $T_\mathrm{eff} < 7000~K$ exhibits ERE, although dust is present with significant optical depths in the nebulae associated with the latter group of stars. Stars with $T_\mathrm{eff} = 7000~K$ have spectral energy distributions (SEDs) with a far-UV cutoff near 170~nm, corresponding to an energy of 7.3~eV, while stars with $T_\mathrm{eff} = 10,000~K$ have their corresponding far-UV flux cutoff near 110~nm, corresponding to photon energies of 11.2~eV \citep{Kurucz74}. The results of \citet{Darbon99} can thus be interpreted to mean that photons with energies within the range between 7.3~eV and 11.2~eV represent the minimum photon energy required to start the ERE process.

The results of \citet{Witt85} are fully consistent with this conclusion. However, the \citet{Darbon99} results eliminate the 217.5~nm absorption peak of the interstellar extinction curve as a possible source of ERE excitation, because $T_\mathrm{eff} = 7000~K$ stellar atmospheres still have significant flux at this wavelength, yet they are not found to be able to excite ERE.

The results discussed above suggest a Stokes shift in excess of 6~eV for the ERE process, a highly unlikely condition for single-step photoluminescence excitation.  Another, more likely possibility is that the far-UV photons simply initiate the ERE by creating and maintaining the emitter,  for example by photo-ionization or -dissociation of a precursor, which could then be capable of photoluminescence upon excitation by abundant lower-energy photons with a more normal Stokes shift.

\subsection{Outline of this Paper}

To decide among these possibilities, it is essential to determine the wavelength of the ERE initiation as accurately as possible.  This must be followed by a comparison of the number density of photons with wavelengths shortward of this limit in a given system with the number density of ERE photons generated within the system.  If the latter is greatly in excess of the former, a two-step initiation/excitation process is the most likely explanation.  The observations reported in this paper provide a basis for a determination of the critical wavelength of ERE initiation/excitation. We employ the fact that the ERE appears in narrow filamentary structures \citep[e.g.][]{Witt89} on the surfaces of molecular cloud clumps in reflection nebulae, in particular in the bright reflection nebula NGC~7023.  We assume that the physical width of a cloud edge as seen in the light of the ERE is determined by the depth of penetration of the photons giving rise to this particular nebular emission. We will compare these penetration depths with those of photons of known wavelengths giving rise to the appearance of the same cloud edges in the light of photo-excited 1-0~S(1)~H$_2$ vibrational fluorescence at 2.12~\micron \citep{Lemaire96, Field98, Takami00, An03} and in the light of scattered radiation in the z-band. In NGC~7023, the former is the result of photo-excitation in the Lyman and Werner bands in H$_2$ \citep{Takami00, Lemaire99}, which occurs near 110~nm and shortward in wavelength, while the latter is the result of simple scattering by dust at the effective wavelength of observation near 900~nm. The penetration depths of the respective exciting radiations are inversely proportional to the respective extinction coefficients for the exciting radiations, with local gas/dust densities canceling when ratios are considered. The unknown wavelength region of ERE initiation can then be estimated from the derived ratios of extinction coefficients, taking into account the opacity sources present in the nebular environment.

Section 2 of this paper presents the observations and reductions; in S ection 3 we present the results. The discussion of the implicationsof our results are contained in Section 4. In particular, we apply the wavelength constraint for ERE initiation to two well-studied ERE sources to assess the likelihood of a two-step excitation process for the ERE. This is followed by a set of conclusions in Section 5. An appendix provides details about the method of calculating the UV/optical absorption spectra of a sample of representative PAH di-cations.

\section{Observations and Data Reduction}
\begin{figure}
\epsscale{0.8}
\plotone{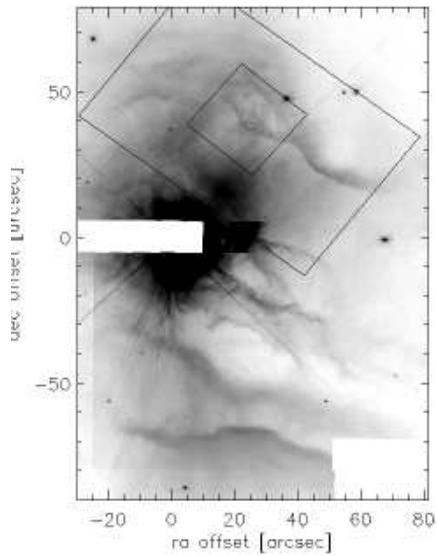}
\caption{Location of the regions shown in Figs.~\ref{fig_acs_images} and
\ref{fig_nic_images} on a F606W WFPC2 image of NGC 7023 \citep{Gordon00}.
\label{fig_big_view} }
\end{figure}

The prominent northwest (NW) filament in NGC~7023 (see
Figure~\ref{fig_big_view}) was imaged with the Advanced Camera for
Surveys (ACS) and the Near Infrared Camera and Multi Object
Spectrometer (NICMOS) on the Hubble Space Telescope (HST) as part of
HST program \#9741.  The ACS images were taken in F475W, F625W,
F850LP, and F656N which correspond to SDSS g, SDSS r, SDSS z, and
H$\alpha$ filters.  The centers of the ACS images were offset to avoid
the bright central star of NGC~7023.  The three broad-band ACS images
were taken to allow for the creation of a continuum-subtracted ERE
image and the narrow band F656N filter to measure the contribution
from scattered H$\alpha$ emission to this ERE image.  The origin of
the scattered H$\alpha$ radiation is the H$\alpha$ emission
line in the spectrum of the central illuminating star, HD 200775. The
NICMOS images were taken with the NIC2 camera and the F212N and F215N
filters which measure the 1-0~S(1)~H$_2$ emission line wavelength
range and associated red continuum at 2.121~$\micron$ and
2.15~$\micron$ respectively.

The observations in each ACS filter were taken split between two
images with a large enough dither to fill in the gap between the two
chips.  The total exposure times were 1000, 1000, 1200, and 1560 s for
the F475W, F625W, F850LP, and F658N filters, respectively.  The two
images per filter were combined with the online multidrizzle and
additionally processed using LACOSMIC \citep{vanDokkum01} to identify
cosmic rays which were not removed with multidrizzle.  Residual cosmic
rays are present in the final mosaics, especially in the gap between
the two ACS chips where we only have a single measurement.  These
residual cosmic rays do not affect our measurements as they rely on
structures much larger than a cosmic ray.

\begin{figure}
\plotone{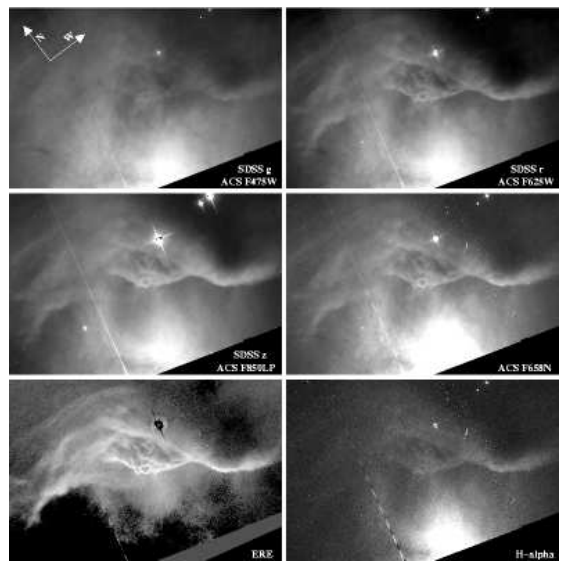}
\caption{A $1\farcm 5 \times 1\farcm 0$ region, rotated $37\fdg 25$ from
North, and centered on  $21^{h} 01^{m} 32\farcs 505, +68\arcdeg 10\arcmin
25\farcs 86$ (2000) is shown for the
four observed ACS filters and continuum-subtracted ERE and H$\alpha$
images.  This region encompasses the NW filament of NGC~7023.
\label{fig_acs_images} }
\end{figure}

A region extracted from the ACS images focusing on the NW
filament is shown in Fig.~\ref{fig_acs_images} for all four observed
filters and the continuum-subtracted ERE and H$\alpha$ images.  The
ERE image was created by first creating a continuum image due to dust
scattered light using a linear combination of the F475W and F850LP
images.  These two filters have central wavelengths (474~nm and
905~nm, respectively) to the blue and red of the ERE peak seen in this
filament which is at $\sim$650~nm \citep{Gordon00}.  The resulting
continuum image was subtracted from the observed F625W image to
produce the ERE image.  The success of the continuum subtraction can
be seen by the much reduced strength of the clump southwest of the
filament (brightest extended source in all four observed ACS bands).
This clump is composed of only scattered light as it has a linearly
decreasing spectrum arcross the three broad ACS bands as measured in a
30x30~pixel box centered on the clump.  The H$\alpha$ image
was created by subtracting the F625W image from the F658N image after
scaling the F625W image by the ratio of fluxes of stars measured in
both images.

The prominence of ERE in this filament is clearly seen as the F625W
image is much sharper than the F475W and F850LP images.  If the F625W
image were dominated by scattered light and not ERE, it would have an
intermediate appearance between the F475W and F850LP.  One possible
reason for this sharper appearance could be H$\alpha$ emission.  This
is not the case as the sharpness of the continuum-subtracted H$\alpha$
image more closely matches the F425W and F850LP images than the F625W
or ERE images.  Note that the H$\alpha$ image has the same morphology
as the scattered-light dominated F425W and F850LP images. This is
because the central star of NGC~7023, HD~200775, has a strong
H$\alpha$ emission line \citep{Gordon00}.  Thus, the
continuum-subtracted H$\alpha$ image traces the scattered H$\alpha$
light of the central star, not H$\alpha$ emission from extended gas in
the nebula.

\begin{figure}
\plotone{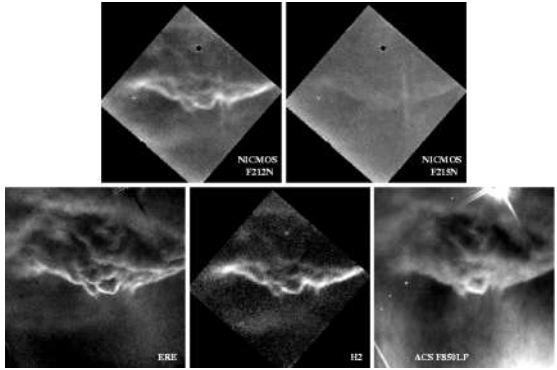}
\caption{A $0\farcm 45 \times 0\farcm 45$ region, rotated $40\fdg 0$ from
North, and centered on $21^{h} 01^{m} 32\farcs 644, +68\arcdeg
10\arcmin 28\farcs 17$ (2000) is shown for the two observed NICMOS
filters, continuum-subtracted ERE, continuum-subtracted H$_2$, and ACS
F850LP images.  This region encompasses the bright portion of the
NW filament of NGC~7023.
\label{fig_nic_images} }
\end{figure}

The observations in the two NICMOS filters were taken split between 4
images with a spiral dither pattern.  The total exposure times were
352 s in both the F212N and F215N filters.  The 4 separate images were
combined using custom scripts to allow for careful masking of bad
pixels and the coronographic spot.  A single star was present in each
image allowing for accurate registration between images and to the ACS
observations.  The mosaics of the F212N and F215N observations are
shown in the top row of Fig.~\ref{fig_nic_images} displayed with the
same scale.  It is clear from these two images that this filament has
very strong H$_2$ emission which was known from existing near-infrared
spectroscopy \citep{Gordon00}.  The H$_2$ image was created by
subtracting the calibrated F215N image from the calibrated F212N
image.  The accuracy of the subtraction of the continuum can be easily
seen as the stellar image disappears completely as well as the
diffraction spikes from the central star of NGC~7023.  In addition to
the NICMOS images, the continuum-subtracted ERE and ACS F850LP images
are shown for the same region in Fig.~\ref{fig_nic_images}.

\section{Results}

\subsection{The ERE, H$_2$, and z-Band Morphology of NW Filaments}

We examined the filamentary morphology of the NW photon-dominated region (PDR) in NGC~7023 on
two spatial scales. The first scale is defined by the more limited
field of view of NICMOS and focuses on the complex, multi-filament
structure in the center part of the NW PDR in NGC~7023 shown in
Figure~\ref{fig_nic_images}. Our observations of the 1-0~S(1) emission
from molecular hydrogen are limited to this field of view, and we are
presenting this image together with the ERE and z-band images limited
to this field. The wider field of view of the ACS defines the second
scale shown in Figure~\ref{fig_acs_images}; it includes the southern and
northern extensions of the filaments in Figure~\ref{fig_nic_images}, for
which we can compare the ERE structure only with the corresponding
structure in the z-band. The data shown in Figures~\ref{fig_acs_images}
and \ref{fig_nic_images} represent the highest-resolution images of the
NGC~7023 NW PDR to date. The location of these images within the
larger extent of NGC~7023 is indicated in Figure~\ref{fig_big_view}.

The structure of the NW PDR shown in the more limited field of NICMOS
(Figure~\ref{fig_nic_images}) is highly complex and consists of a
series of narrow filaments with typical widths of 1 arcsec when viewed
in the light of ERE and the light of H$_2$. These filaments are fully
resolved in both images. The angular resolution is marginally higher
in the ACS images compared to the NICMOS images. This is based on
comparisons of stellar images which yielded ratios
FWHM(ACS)/FWHM(NICMOS) = 0.93, with stellar images having diameters
with FWHM of 0.1025 and 0.1102 arc seconds, respectively.  When viewed
in the z-band, which is dominated by dust-scattered light, the same
structures appear much more diffuse, although they were imaged with
the same angular resolution. This comparison suggests that the
sharpness of the structures as seen in ERE and H$_2$ is not a result
of a limited physical extent of the structures, e.g. thin sheets
viewed edge-on, but that they have a physical width that is wider than
indicated by the ERE and H$_2$ filament images. The narrow ERE and
H$_2$ filaments, therefore, may represent just the edges of more
extended molecular cloud clumps illuminated by the central B3Ve star
in NGC~7023, HD~200775.

\begin{figure}
\plotone{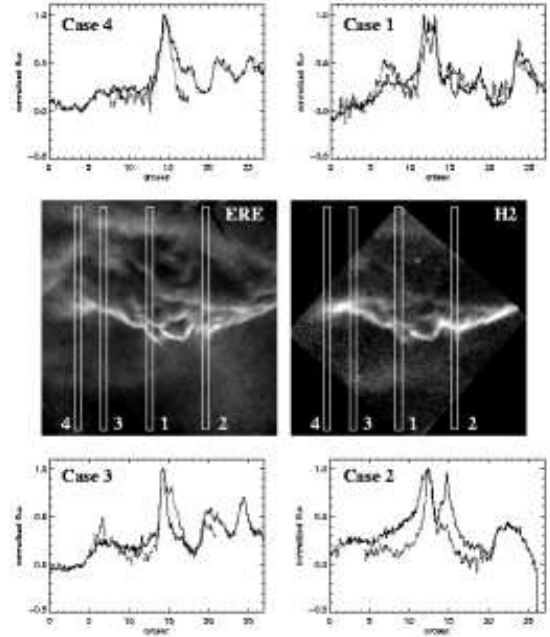}
\caption{Four cuts through the ERE (solid line) and H$_2$ (dotted
line) images are given
illustrating the 4 cases seen.  These cases are 1) near-perfect match
between ERE and H$_2$ filametns, 2) evidence for H$_2$ self-shielding,
3) H$_2$ filaments broader than ERE, and 4) H$_2$ filaments narrower
and in front of ERE filaments. The x-axes of the graphs increase in
the direction of propagation of the exciting radiation.
\label{fig_ere_h2_cuts} }
\end{figure}

There is independent evidence from the interferometric HCO$^+$
observations by \citet{Fuente96} for the existence of at least two but
possibly four high-density molecular clumps in our field of view that
can be distinguished on the basis of their differing radial
velocities. In particular, the extended filament that begins on the
right-hand side of Figure~\ref{fig_acs_images} and turns upward, passing
near the embedded star \citep[star K in the map of][]{An03}, appears to
be a coherent structure moving at 2.4~km~s$^{-1}$, while the group of
narrow filaments shown in Figure~\ref{fig_nic_images} closest to the
illuminating source belongs mostly to a structure moving at
4.0~km~s$^{-1}$. The apparent complexity of the filamentary structure
of the NW PDR is most likely the result of the line-of-sight
superposition of several independently moving molecular clouds, whose
surfaces facing HD~200775 are currently being photo-dissociated.

A comparison of the morphology of the H$_2$ and ERE filaments in
Figure~\ref{fig_nic_images} reveals two significant facts. One, the
ERE filaments appear to be about as sharp as the corresponding H$_2$
filaments. We will investigate this aspect further in
Section~\ref{sec_ere_width}; for now we can conclude that the
radiation initiating the H$_2$ fluorescence and the ERE, respectively,
apparently faces similar optical depths per unit mass of molecular
cloud environment. Two, the spatial correlation between the
distribution of the ERE and the H$_2$ emission in
Figure~\ref{fig_nic_images} is not exceptionally strong. ERE is
present wherever H$_2$ emission is seen, but there are bright ERE
filaments, especially in the upper part of
Figure~\ref{fig_nic_images}, for which the corresponding H$_2$
features are either much weaker in intensity or absent altogether. A similar
qualitative relationship between ERE and H$_2$ emission has previously
been noted by \citet{Field94} in a study of the reflection nebula
NGC~2023 and was confirmed in a more detailed study of NGC~7023 by
\citet{Lemaire96}.

When comparing the detailed profile shapes and positions of ERE and
H$_2$ filaments, we can identify four classes, illustrated in
Figure~\ref{fig_ere_h2_cuts}. These can be described as follows. Case
1 represents locations where ERE and H$_2$ exhibit nearly identical
structures, agreeing both in width and relative surface brightness. Out of 36 cuts examined, seven fall into this
category. Class 2 includes cases (17/36) where the first H$_2$
filament is very narrow and recessed with respect to the corresponding
ERE filament. A second ERE filament is not matched by H$_2$ in terms of its surface brightness. We
believe these to be instances of H$_2$-self-shielding. Case 3
represents instances (9/36) where the H$_2$ filament is broader than
the corresponding ERE filament, while Case 4 illustrates the rather
rare (3/36) circumstance where the H$_2$ filament is narrower than the
ERE filament and peaks slightly in front of the ERE filament.

The near-absence of H$_2$ emission (Case 2) in regions exhibiting
bright ERE cannot be explained by differences in line-of-sight
extinction. We just concluded that both types of emission appear to be
initiated by photons of similar energy; the resulting outgoing
1-0~S(1) H$_2$ emission faces an opacity only 1/8 as large as the
resulting ERE, the former occurring in the near-IR. Hence, if one type of emission should be missing due to
extinction, it would be the ERE, not the H$_2$ emission, contrary to
our observations. As noted by \citet{Field94} and \citet{Lemaire96},
this result provides a strong argument against models, in which the
ERE carrier is initiated chemically by hydrogenation of carbonaceous
dust in the hot atomic hydrogen gas found in H$_2$ photo-dissociation
fronts \citep{Witt88, Duley90}. Lack of carriers also cannot be the
explanation for the absence of H$_2$ luminescence; H$_2$ molecules are
the principal constituents of molecular clouds in the ISM. If H$_2$ luminescence is weak or absent in a dense molecular cloud, H$_2$ molecules are not sufficiently excited.
As a result, H$_2$ remains molecular and the hot HI atoms required for hydrogenation of carbonaceous dust are not available. If ERE is observed to be bright in such regions, its origin must be in processes other than hydrogenation.

\begin{figure}
\plotone{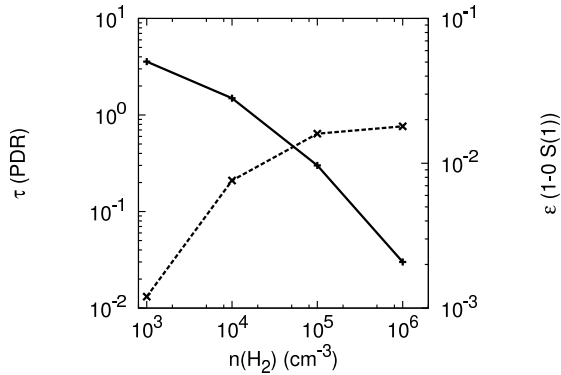}
\caption{The density-dependence of the dust optical depth of a PDR
(solid line) and the efficiency of converting far-UV photons into
near-IR 1-0~S(1) H$_2$ photons (dotted line) is shown. The NGC 7023 NW
PDR involves molecular gas with densities ranging from n(H$_2$) $\geq$
10$^4$ cm$^{-3}$ to n(H$_2$) $\geq$ 10$^6$ cm$^{-3}$ \citep{An03}.
\label{fig_ere_n_tpdr} }
\end{figure}

We notice that ERE filaments in the NW PDR that lack H$_2$
counterparts of corresponding surface brightness tend to be broader
than those that have bright H$_2$ counterparts. If this broader, more
diffuse appearance is a reflection of the penetration depth of the
radiation initiating the ERE, this then indicates that the density in
the ERE filaments lacking H$_2$ counterparts is substantially
lower. Under these conditions, the UV pumping efficiency as measured
by the number of 1-0~S(1) photons emitted per incident photon in the
wavelength range 110.8-91.2~nm is significantly reduced
\citep{Draine96}, because dust is absorbing an increasing fraction of
the photons needed for the excitation of H$_2$. This is because
molecular hydrogen is unable to be re-formed on the surfaces of dust grains as quickly under lower-density conditions, the rate being proportional to the product of the dust density and the density of atomic hydrogen. As we had seen earlier, the ERE carrier is part of the
dust and a likely competitor for photons in the far-UV range. Thus,
with competition lacking from molecular hydrogen, the ERE carrier is
able to absorb a greater fraction of the available photons under
lower-density conditions. We illustrate the dependence of the dust
optical depth of a PDR and the efficiency of converting absorbed
far-UV photons into near-IR photons in the 1-0~S(1) H$_2$ transition
in Figure~\ref{fig_ere_n_tpdr}, calculated from relations provided by
\citet{Draine96}. The dust optical depth of the PDR is measured from
the front of the molecular cloud to the point at which 50\% of the
H$_2$ is photo-dissociated.

It appears likely, therefore, that the lack of a detailed spatial
correlation between ERE filaments and the 1-0~S(1) emission of H$_2$ is
a result of density variations by about one order of magnitude between
high-density filaments, where both ERE and H$_2$ are bright, and lower
density cloud faces, where the ERE is dominant.

\subsection{Determination of R$_V$ in NW Filaments}

When viewing the filaments in the light of different emission
mechanisms, e.g.\ the fluorescent 1-0~S(1) vibrational emission of H$_2$
at 2.12~\micron\ and the much more diffuse appearance of the same
structures in the light of dust-scattered light in the z-band
(Figure~\ref{fig_nic_images}), we are
comparing the relative penetration into the cloud surfaces of the
respective exciting radiations. In the case of the fluorescent H$_2$
emission, the exciting radiation is Lyman and Werner band photons
absorbed by H$_2$ molecules at wavelengths $\lambda \leq 110.8$~nm. The
opacity restricting the transfer of such photons into the cloud faces
is most certainly extinction by dust and, depending on local density
conditions, self-shielding by H$_2$ \citep{Draine96}. In the
z-band, dust extinction is the only significant opacity source. The
relative penetration depth along a given line-of-sight from the
central star into a molecular cloud is determined by the inverse ratio
of these two opacities. Taking into account the possible contribution
by H$_2$ self-shielding to the opacity, the ratio of the widths of the
H$_2$ and the z filaments at identical locations, then, is an upper
limit to the ratio of the dust opacities at the two effective
wavelengths involved, 110.8~nm and 900~nm, respectively. This ratio
can now be used to identify the appropriate dust extinction curve for
the molecular cloud fronts in the NGC~7023 NW PDR.

\begin{figure}
\plotone{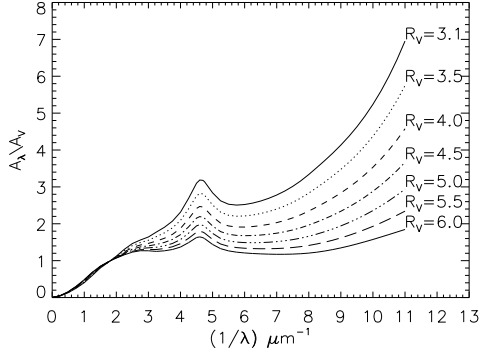}
\caption{Extinction curves generated from the CCM R$_V$-dependent
relationship are shown for a range of R$_V$ values.
\label{fig_ccm} }
\end{figure}
\begin{figure}
\plotone{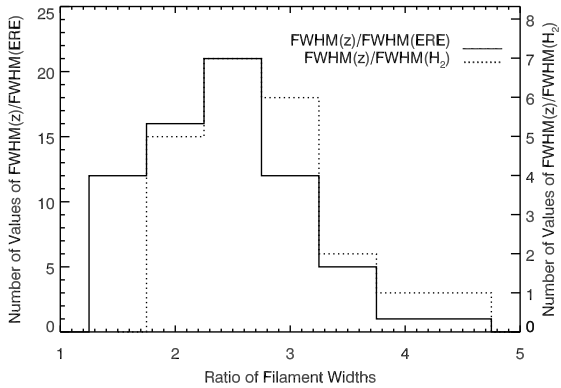}
\caption{Histogram of the ratios FWHM(z)/FWHM(H$_2$) and
FWHM(z)/FWHM(ERE) of filaments widths.
\label{fig_z_h2_histo} }
\end{figure}

\begin{figure}
\plotone{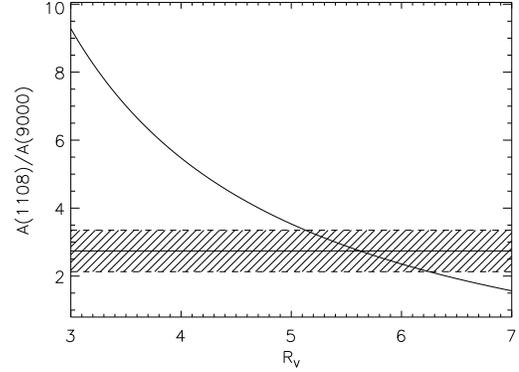}
\caption{The ratio of extinctions at 110.8~nm and 900~nm is plotted as
a solid line.  The allowed region from the measurement of the
FWHM(z)/FWHM(H$_2$) ratio is shown as the shaded region.
\label{fig_a1104a9000} }
\end{figure}

\citet[hereafter CCM]{Cardelli89} showed that the wide range of
extinction curves encountered in different galactic environments can
be reproduced as a family of functions dependent upon a single
parameter, the ratio of total-to-selective extinction, R$_V$, as
illustrated in Figure~\ref{fig_ccm}. The low-density diffuse ISM in
the Milky Way is most commonly represented by CCM curves with R$_V =
3.1$ while dense molecular cloud material exhibits CCM extinction
curves with R$_V$ in the range from 5 to 6. As long as we adopt the
CCM formalism, these curves can be distinguished from each other by
measuring a single ratio of extinctions at two widely separated
wavelengths. As we noted before, such extinction ratios can be
estimated from the relative penetration depth at the respective
wavelengths. We measured the FWHM of the H$_2$ and z-band filaments
(Figure~\ref{fig_nic_images}) at 22 positions where such measurements
could be carried out without interference from blends by closely
spaced H$_2$ filaments, whose z-counterparts would merge into a single
profile. A histogram of the measurements of the ratio of
FWHM(z)/FWHM(H$_2$) is shown in Figure~\ref{fig_z_h2_histo} by a
dotted line. The average of the measured ratios of FWHM(z)/FWHW(H$_2$)
is $2.74 \pm 0.61$. In Figure~\ref{fig_a1104a9000} we show the ratio
of the dust extinction cross sections at 110.8~nm and 900~nm, the
latter being the effective wavelength of the z-band, as a function of
R$_V$ for CCM extinction curves, and we have entered our measured
ratio as a horizontal bar. We conclude from this figure that a CCM
extinction curve with R$_V = 5.62^{+0.61}_{-0.48}$ is a good estimate
for the dust extinction curve in the molecular clouds making up the NW
PDR in NGC~7023. If H$_2$ self-shielding is a significant contributor
to the total extinction for the H$_2$-exciting radiation, this value
of R$_V$ is a lower limit. The fact that R$_V = 5.6$ is very typical of dust 
in molecular cloud environments (R$_V = 5.5$) suggests, however, that
H$_2$ self-shielding is a minor contributor to the opacity for the
H$_2$-exciting radiation in the filaments measured in this experiment.

\subsection{Width of the ERE Filaments}
\label{sec_ere_width}
\label{sec_ere_measure}

Employing the same rationale as in the previous section, we now
compared the relative widths of the ERE and z filaments in
Figures~\ref{fig_acs_images} and \ref{fig_nic_images} in identical
locations. A total of 41 measurements were made in the larger field
(Figure~\ref{fig_acs_images}) and 26 such measurements were made in
the smaller field (Figure~\ref{fig_nic_images}). A histogram of the
resulting ratios FWHM(z)/FWHM(ERE) is shown in
Figure~\ref{fig_z_h2_histo} (solid line). The two sets of measurements
resulted in average ratios of FWHM(z)/FWHM(ERE) of $2.52 \pm 0.65$ and
$2.32 \pm 0.64$, respectively, with a weighted average of $2.40 \pm
0.65$. Comparing this value to the corresponding ratio of
FWHM(z)/FWHM(H$_2$) $= 2.74 \pm 0.61$, we conclude that on average the
ERE filaments are about 10\% broader than the corresponding H$_2$
filaments, if they are present, and that the opacity facing the
radiation that initiates the ERE is on average, therefore, about 10 \% less than
the opacity encountered by the radiation that excites the
H$_2$~1-0~S(1) radiation.

\subsection{Wavelength of ERE Initiation}
\label{sec_ere_wave}

\begin{figure}[tbp]
\plotone{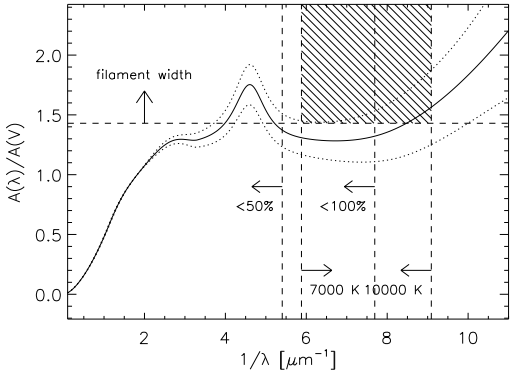}
\caption{The limits on the ERE excitation wavelength are shown on top
of a CCM R$_V$ = 5.62 extinction curve. The constraint imposed by the
ERE filament widths is discussed in Section~\ref{sec_ere_measure}; the
constraints due to the limits on the effective temperatures of the
exciting stars are discussed in Section~\ref{sec_ere_wave}; and the
two vertical lines labeled $<$ 100\% and $<$ 50\% related to the
insufficiency of far-UV photons for the actual excitation of the ERE
are discussed in Section 4.4.
\label{fig_ere_limits} }
\end{figure}

In order to identify the wavelengths of the radiation that appears to
be required for ERE initiation, we plotted a CCM extinction curve for
R$_V = 5.62$, plus two dotted extinction curves representing the
uncertainty limits in the R$_V$ value, found to be appropriate for the
dust in the molecular material in NGC 7023, in
Figure~\ref{fig_ere_limits}. Assuming a ratio A(z)/A(V)~=~0.59 for the
R$_V = 5.62$ case, we place our measured ratio of FWHM(z)/FWHM(ERE)
onto this graph as a horizontal line at A($\lambda$)/A(V)~=~1.43. This
constraint implies that only those wavelength ranges in which the
extinction $\emph{exceeds}$ this horizontal limit exhibit adequate
dust opacity to be consistent with our data in
Figure~\ref{fig_z_h2_histo}. Resulting from the multi-valued nature of
the extinction curve, this identifies two spectral regions: the range
from 250~nm to 189~nm, where the horizontal line crosses the 217.5~nm
bump in the extinction curve, and the region shortward of 118~nm in
the far-UV, where the far-UV rise of the extinction curve reaches
values in excess of our opacity constraint. Thus, similar to the
investigation by \citet{Witt85}, our method alone does not result in
the identification of a unique spectral region responsible for ERE
initiation, but it clearly excludes near-UV or optical radiation shortward of the ERE band from consideration.

We can now narrow down the choice between the two UV spectral ranges
identified above by applying the results of \citet{Darbon99} as a
further constraint. In short, \citet{Darbon99} found that stars with
$T_\mathrm{eff} \geq 10,000 K$ can excite ERE and very commonly do, while
stars with $T_\mathrm{eff} \leq 7000 K$ do not. We entered the FUV cut-offs
of the SEDs of atmospheres with these two temperatures in
Figure~\ref{fig_ere_limits} as two vertical lines. The long-wavelength
limit of the spectral region in which ERE is initiated radiatively is
defined by the point at which our applicable extinction curve enters
from below the cross-hatched area limited by the three constraint
lines. For R$_V = 5.62$, this wavelength is 118~nm, corresponding to a
photon energy near 10.5~eV. Therefore, the application of the
constraints by \citet{Darbon99} eliminates the 217.5~nm extinction
band as the source of excitation of the ERE. Our new constraint based
upon the penetration of the exciting radiation into the cloud face
considerably narrows the acceptable wavelength range for ERE
initiation allowed by the \citet{Darbon99} results, however. Both sets
of constraints are consistent with the possibility that photons with
wavelengths between 118~nm and 91.2~nm, the latter limit being set by the competing absorption by HI, can contribute to ERE initiation.

\section{Discussion}

Having obtained a fairly accurate estimate of the wavelengths of the
photons needed to initiate the ERE will be critical for determining
further details of the ERE process as well as the identity of the ERE carrier. As we
stated earlier, our present observations do not allow us to determine
whether photons with $E > 10.5$~eV are required to excite the ERE
carrier, with (at best) one ERE photon resulting from each absorption
of each far-UV photon, or whether the presence of far-UV photons is
merely required for the creation of the ERE carrier, e.g.\ by either
photo-ionization or photo-dissociation of an ERE-precursor, followed
by pumping of the newly created ERE carrier by lower-energy optical
photons. However, this issue can be resolved directly by counting the
number of photons available in the 118 nm to 91.2 nm range that are
absorbed in a given system and comparing this to the number of ERE
photons emitted. We will return to this question in Sections
\ref{sec_red_rect} and \ref{sec_cirrus}.

\subsection{Impact on Existing ERE Models}

While the nature of the ERE as a dust-related interstellar
photoluminescence process has been undisputed, many diverse proposals
have been advanced for possible identifications of the ERE carrier
\citep[see][for a review]{Witt04}. With the determination of the
minimum wavelength of the radiation required for ERE initiation, many
of these proposals can now be ruled out. In discussing these
proposals below, we will focus primarily on the excitation
requirements and not on respective failures to meet other
observational constraints.

\subsubsection{Neutral PAH Molecules}
\label{sec_pah}

\citet{dHendecourt86} proposed that fluorescence, i.e.\ electronic
transitions from the S$_1$ state to vibrational levels of the S$_0$
ground state, or phosphorescence, i.e. electronic transitions from the
T$_1$ state to vibrational levels of the S$_0$ ground state, in
neutral polycyclic aromatic hydrocarbons (PAHs) could be responsible
for the ERE. The same species are considered the likely source of the
ubiquitous mid-IR aromatic emission features (AEF) \citep{Bakes04,
Onaka04, Peeters04}. In the laboratory, many PAH molecules are known
to fluoresce efficiently at optical wavelengths. However, as noted by
\citet{dHendecourt86}, PAH fluorescence generally occurs at much
shorter wavelengths than the ERE band while phosphorescence frequently
does occur in the same spectral region as the ERE. Although the T$_1$ -
S$_0$ transitions are spin-forbidden, isolated PAH molecules in the ISM
could conceivably be strong sources of phosphorescence, provided that
the T$_1$ state can attain high populations. The T$_1$ state, which is
always at a lower energy compared to the S$_1$ state in neutral PAH
molecules, is populated by intersystem crossings from excited states
of the singlet manifold to highly-excited vibrational levels of the
T$_1$ state. This implies, therefore, that the excitation of
phosphorescence occurs via the same absorption bands that excite the
fluorescence in these molecules. The absorption spectra of neutral PAH
molecules are well known; their absorption bands are found immediately
shortward in wavelength of where their S$_1$ - S$_0$ fluorescence
occurs, with the strongest bands found generally near 200~nm. Thus,
the wavelengths of radiation required for the excitation of PAH
fluorescence and PAH phosphorescence are two to three times longer
than what was found in the present investigation. Recently, neutral
PAH fluorescence \citep{Vijh04, Vijh05} has been detected in the Red
Rectangle, an object in which ERE is present as well, and the spatial
distributions of the two emissions were found to be distinctly
different. This is only further support for our conclusion that
neutral PAH phosphorescence from these same molecules is not the source of ERE.
Very large neutral PAH molecules can fluoresce in the red spectral range, although their efficiency is low \citep{Herod96}. Even if their quantum yield were higher, their dominant absorption bands are located in the optical range, which is in direct conflict with our measurements, requiring far-UV photons for ERE initiation.

\subsubsection{PAH Clusters}

\citet{Seahra99} suggested that PAH clusters in the form of stacks or
aggregates of PAH molecules with up to 700 carbon atoms could produce
ERE-like photoluminescence in a band centered near 700 nm. The same
authors \citep{Duley98} had previously proposed that the same PAH
clusters could be responsible for the interstellar absorption band at
217.5~nm as well as the mid-IR AEF bands. The energy for both the ERE
and the AEF in this model was to come from the absorption in the
217.5~nm band. While the present penetration study does not exclude
this possibility, the application of the \citet{Darbon99} constraint
(Figure~\ref{fig_ere_limits}) does. The PAH cluster model is therefore
not supported by the ERE excitation results.

The PAH cluster model of \citet{Seahra99} also makes other predictions that
are in conflict with observations. In particular, it predicts the presence of two side bands
to the main ERE band, appearing at wavelengths near 0.5 and 1.0 $\mu$m, neither of which has been found in 
observational data \citep{Gordon00}. Furthermore, it predicts a near-constant peak wavelength for the main ERE band 
near 0.7 $\mu$m, while the observed ERE peak shifts by more than 200~nm in response to changes in the radiation envioronment, in which the ERE is being produced \citep{Smith02}

\subsubsection{Hydrogenated Amorphous Carbon}

Photoluminescence by grains consisting of hydrogenated amorphous
carbon (HAC) or coated by HAC mantles was an early candidate for the
ERE process \citep{Duley85, Witt88, Duley90, Jones90}. With a band gap
of over 3 eV \citep{Robertson96}, HAC not only exhibits efficient
photoluminescence at wavelengths much shorter than observed in ERE, it
also is efficiently excited at optical/near-UV wavelengths
\citep{Watanabe82}, which is in conflict with our present results. The
latter objection can also be raised against a material closely related
to HAC, referred to as quenched carbonaceous composite (QCC)
\citep{Sakata92}. While the band gap of various HAC materials can be
reduced by a variety of treatments, the most efficient excitation
occurs always just shortward of the emission band resulting from HAC
photoluminescence. This puts these models in serious conflict with our
new excitation constraints.

\subsubsection{Silicon Nanoparticles}

An ERE model that received intensive study in recent years involves
photoluminescence by oxygen-passivated silicon nanoparticles (SNPs),
containing between about 200 and 6000 silicon atoms \citep{Witt98,
Ledoux98, Ledoux00, Ledoux01, Ledoux02, Smith02, Li02}. The band gap
of SNPs is size dependent due to quantum confinement and with suitable
limits on the nanoparticle sizes it readily matches the observed
wavelengths of the ERE band. However, again as in all instances of
typical semiconductor photoluminescence, the most effective excitation
occurs at energies just slightly above the band gap energies, i.e.\
slightly above 2.3~eV. In addition, SNPs are photo-ionized when
exposed to photons with energies in excess of 5.1~eV \citep{Fuke93},
with the result that photoluminescence from subsequent excitations is
quenched \citep{Nirmal99, Smith02}. Consequently, our new constraints
regarding ERE initialization also do not favor the SNP model.

\subsection{Possible New ERE Carriers}

All previously considered ERE carrier models appear to be inconsistent
with the requirement that photons with energies in excess of 10.5~eV
are required to initiate the ERE in astrophysical environments. There
are several reasons for assuming that the energy limit of 10.5~eV is
not directly related to the actual excitation of the ERE.  Below, we
will argue that the number density of photons with energies in excess
of 10.5~eV in typical ERE-producing environments is insufficient to
account for the number of ERE photons emitted from those environments,
if the $E > 10.5$~eV photons were to be the direct source of excitation,
even if a photon conversion efficiency of 100\% is assumed.  A
possible solution to this dilemma is provided by models in which the
$E > 10.5$~eV photons simply create the ERE carrier, with a small number
of FUV photons required for its maintenance. The ERE carrier thus
created needs to have the ability for efficient photoluminescence in
the 540-900~nm wavelength range and strong absorption throughout the
optical/UV regions of the spectrum, where ample photons for the
pumping of the ERE are available. Furthermore, the carrier species
involved in such a process must be abundant in the ISM in order to
account for the contribution of the ERE carrier to the absorption in
the ISM, amounting to about 10\% of the photons of the Galactic
interstellar radiation field in the 540-91.2~nm wavelength range
\citep{Gordon98}.

\begin{figure}[tbp]
\epsscale{0.8}
\plotone{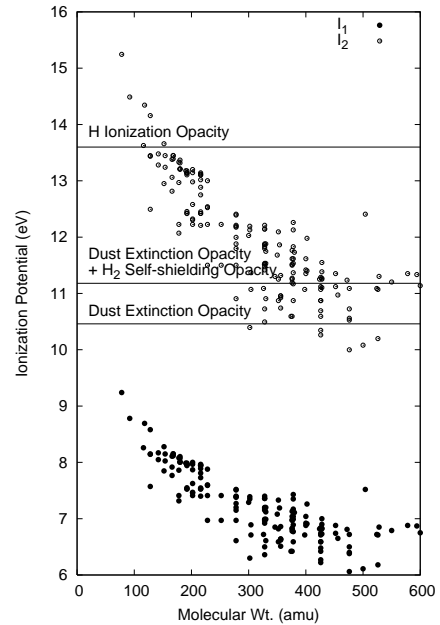}
\caption{First ionization potentials of polycyclic aromatic
hydrocarbon molecules (filled circles) \citep{Eilfeld81} and their
corresponding cations (open circles) \citep{Leach96} versus molecular
weights. The three horizontal lines correspond to the cutoff of the
far-UV radiation field due to H-ionization opacity (top), the onset of
the energy regime in which dust extinction and H$_2$ self-shielding
opacity compete for photons (middle), and the lower limit on the
energy for ERE initiation (lowest), respectively.
\label{fig_ev_vs_amu} }
\end{figure}

In terms of abundance and total cross section, interstellar PAHs meet
these requirements. However, as shown in Section~\ref{sec_pah},
neutral PAHs cannot account for the observed ERE. By contrast, doubly
ionized PAH molecules offer interesting prospects. In
Figure~\ref{fig_ev_vs_amu} we have plotted the first ionization
potentials of neutral PAH molecules and of PAH mono-cations as a
function of the size of the molecules as measured by their molecular
weights. We note that most PAHs with up to 50 carbon atoms can be
readily ionized into the mono-cation stage with photons with energies
in the 6~eV to 8~eV range, and can subsequently be sent into the
di-cation state with photons in the energy range from 10.5 to 13.6~eV.
It should, therefore, be possible to find PAHs in the di-cation state in
environments characterized by the presence of neutral or molecular
hydrogen, as previously suggested by \citet{Leach87, Leach96}.  PAH
di-cations are closed-shell systems similar to neutral PAH molecules
with S$_1$-S$_0$ transitions occurring at optical wavelengths longer
than the corresponding transitions in neutral molecules. Fluorescence
could thus be expected in the ERE range. The viability of such a model
requires laboratory spectroscopy on PAH di-cations that could reveal
the presence of optical fluorescence, following the excitation through
near-UV/optical absorptions. No such experimental data exist at this
time.

\begin{figure}[tbp]
\plotone{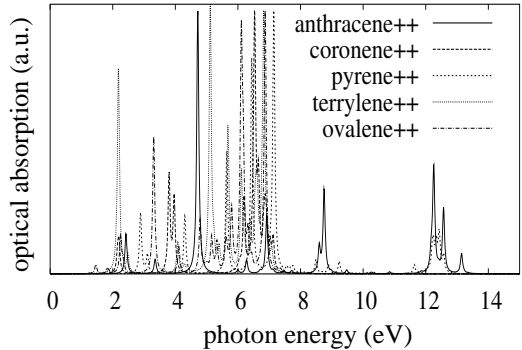}
\caption{Model spectra of representative PAH di-cations.
\label{fig_model_pah} }
\end{figure}

We can, however, obtain a general idea of the expected
spectroscopic properties of PAH di-cations, using density-functional
theory modeling calculations (TD-DFT) \citep{tddft}.  This theory
incorporates electronic screening and relevant correlation effects
\citep{excited} for vertical  electronic excitations in  the ground-state
geometry and thus represents a fully {\sl ab initio} formalism for
computing excited states.  Over the past decade, TD-DFT has given
promising results for finite systems (atoms, molecules, clusters)
\citep{excited,tddft,basis}.  In this paper, we use the TD-DFT
implemented in the Gaussian 03 package \citep{gaussian}. Based on the
ground-state structures for all clusters, which were optimized by
using DFT \citep{dft} with the nonlocal hybrid B3LYP functional
\citep{b3lyp}, we compute the excited energies of both singlet and
triplet states \citep{lewis}, the oscillator strengths, and the
optical absorption gap (generally defined by the energy of the first
dipole-allowed transition for finite systems
\citep{excited}). To ensure the accuracy of our calculations, we
choose the 6-31G(d,p) Pople-type basis set which adds polarization
functions to the atoms \citep{basis}.  

In Figure~\ref{fig_model_pah}
we show the composite absorption spectra of five PAH di-cations,
including results for pyrene++ and anthracene ++, whose likely
presence in their neutral and singly-ionized form in the Red Rectangle (HD 44179) has been demonstrated by both
their blue fluorescence as well as their spectroscopic signature of
the first ionization \citep{Vijh04, Vijh05}. When compared with
available experimental data for neutral and once-ionized PAH species,
TD-DFT calculations agree remarkably well with experimental
measurements \citep[][and references therein]{Malloci04}. The spectra in Figure~\ref{fig_model_pah} show that
representative PAH di-cations absorb strongly in the energy range from
2~eV to 7.5~eV, suggesting a characteristic band gap of
2~eV. Photoluminescence across this band gap could be responsible for
the ERE.

 A still controversial issue is the question of the stability
of PAH di-cations against Coulomb explosion. A recent study of 
the fragmentation
pathways of the di-cation of benzene by \citet{Rosi04} revealed an
unexpectedly high degree of stability. These authors suggest that the
di-cations of multi-ringed PAHs could exhibit a still higher degree of
stability and thus play a substantial role in
interstellar chemistry and photophysics. Experimental results reported by 
\citet{Leding99}, involving the production of di-cations of the aromatic
species benzene, monodeuterated benzene, toluene, and naphthalene
through multi-photon soft ionization, have shown that the resultant di-cations are 
in fact stable. Multi-photon soft ionization is equivalent to interstellar conditions in the sense
that the di-cations are produced with only a small amount of excess vibrational energy.
In the ISM, this excess energy would have an upper limit equal to the
difference between the energy of the Lyman limit of atomic hydrogen (13.6 eV)
and the second ionization potential of an individual PAH molecule. In a PDR, such as the ERE filaments
studied in this paper, photons with energies in excess of 11.14 eV will be absorbed preferentially by
molecular hydrogen. Thus, to the extent that second ionizations of PAHs are possible with photons in the energy range from 10.5 eV - 11.14 eV, these would
be extremely soft indeed.Therefore, both theory and experiment provide strong encouragement for considering the likely presence of
PAH di-cations in interstellar environments where ERE is observed.

\subsection{Consistency Check: The Red Rectangle}
\label{sec_red_rect}

The two principal results emerging from the study of the excitation of
the ERE in the NGC~7023 NW PDR are: (1) the ERE initiation requires
photons with energies $> 10.5$~eV, e.g.\ for the creation of the ERE
carrier; (2) the ERE process is most likely a two-step process, in
which the first step creates the carrier via ionization or
dissociation, requiring in excess of 10.5~eV, and the second step
produces the actual excitation of the ERE carrier via
longer-wavelength photons. In this section we will examine the Red
Rectangle, one of the two environments where ERE is being produced
with very high efficiency, the other being the high-$|b|$ Galactic
cirrus.

Two aspects are of greatest significance, the morphology of ERE
sources and the ERE energetics. The Red Rectangle is a well-studied
bi-polar protoplanetary nebula with a circum-stellar disk obscuring
the central source \citep{Menshchikov02, Cohen04}. The central source
consists of a $L \sim 6000 L_{\sun}$ AGB star with a $T_\mathrm{eff} \sim
8250$~K atmosphere \citep{Vijh05} and a low-mass ($M \approx 0.35
M_{\sun}$) hot white dwarf of unknown luminosity, which could amount
to at most $\sim 100 L_{\sun}$ \citep{Menshchikov02, Driebe98}. This
upper limit to the the white dwarf's luminosity would apply only,
however, during the $\sim 10^5$~yr duration of the stellar core's
contraction from the AGB region of the Hertzsprung-Russell diagram to
the top of the white dwarf cooling sequence. Once the stellar core has
reached a size corresponding to the white dwarf mass-radius relation,
a more typical luminosity of a $T_\mathrm{eff} = 60,000$~K hot white dwarf is
about $6.5 L_{\sun}$. The relevance of these luminosity limits will
become apparent presently.

Spatially, the ERE in the Red Rectangle is confined to geometrically
thin regions within the walls of the bi-conical outflow cavities
\citep{Schmidt91, Cohen04}, leading to the
appearance of the X-shaped bipolar structure, while scattering in the
500 to 350 nm wavelength range produces a spherical reflection nebula
devoid of the bipolar structure \citep{Cohen04}. The confinement of
the ERE to the walls of the outflow cavity is in agreement with our
finding of an initiation of the ERE by high-energy UV photons. These
photons, stemming predominantly from the hot white dwarf, are
spatially confined to the outflow cones and its walls by the optically
thick circum-binary disk and the steeply rising far-UV opacity of the
nebular dust in the Red Rectangle \citep{Vijh05}. At wavelengths
longward of 170~nm ($E < 7.3$~eV), the nebular opacity is greatly
reduced. Thus, the mid-UV, near-UV and optical photons, which are
produced almost entirely by the cooler but extremely luminous AGB
star, are able to penetrate the walls of the outflow cavity and
produce the nearly spherical blue reflection nebula component of the
Red Rectangle \citep{Cohen04}. They are also able to efficiently pump
the ERE carriers produced in the thin wall regions of the outflow
cones.

The ERE energetics of the Red Rectangle is particularly useful in
supporting the concept of a two-step process of ERE excitation.
\citet{Schmidt80} reported the band-integrated ERE luminosity of the
entire Red Rectangle nebula as $L_{ERE} \approx 0.6 (D/280~pc)^2
L_{\sun}$. Using the currently accepted distance of 710~pc
\citep{Menshchikov02, Hobbs04}, we arrive at an estimate of $L_{ERE}
\approx 3.9 L_{\sun}$ for the Red Rectangle. This value is uncorrected
for attenuation by the optically thick circum-binary disk surrounding
the stellar sources as well as for any interstellar extinction along
the line of sight.  Given the unusual color characteristics of the RR
and the non-isotropic morphology of the RR, these attenuations are
difficult to estimate. The light of the central AGB star alone suffers
an attenuation in excess of $A_V$ = 4 magnitudes \citep{Vijh05}.  We,
therefore, consider it a conservative estimate that the total ERE in
the RR suffers an attenuation by about 1 magnitude, resulting in a
total estimated ERE luminosity of $L_{ERE} \approx 10 L_{\sun}$.
If photons with $E > 10.5$~eV were responsible for the actual
excitation of the ERE in a single-step process, and if the excitation
process were 100\% efficient, with one ERE photon of 1.9~eV produced
for every absorption of a $E > 10.5$~eV photon, the total luminosity
in $E > 10.5$~eV photons required for this excitation would be
$L_{FUV} > 55 L_{\sun}$. Thus, with somewhat more realistic
assumptions about the ERE excitation efficiency of possibly 30\% to
50\%, we would require in excess of 100 $L_{\sun}$ in far-UV
photons to produce the observed ERE luminosity of the Red Rectangle,
well in excess of what is likely available from the hot
white dwarf companion.

If, by contrast, the far-UV photons emitted by the hot white dwarf
companion of the central AGB star are used solely for creating and
maintaining the ERE emitters, the AGB star ($T_\mathrm{eff} \approx 8250 K$;
$L \approx 6000 L_{\sun}$) has more than enough energy to excite $L_{ERE}
\approx 10 L_{\sun}$ with mid-UV and optical photons. Assuming a Stokes
shift of 1~eV typical for photoluminescence, absorption of an
excitation energy of less than $15 L_{\sun}$ would suffice to generate
$L_{ERE} \approx 10 L_{\sun}$, assuming 100\% efficiency.
\citet{Gordon98} have estimated the ERE photon conversion efficiency
to be $> 10$\%, which would lead to an upper limit for the ERE
generating luminosity of $150 L_{\sun}$. This is still only about
2.5\% of the luminosity of the AGB star.  Thus, a two-step process of
ERE excitation is entirely feasible, given the energetics of the Red
Rectangle nebula system.

\subsection{Consistency Check: The High-$|b|$ Galactic Cirrus}
\label{sec_cirrus}

Next to the Red Rectangle, the high-$|b|$ Galactic cirrus is another
environment in which the intensity of the ERE is roughly equal to that
of the dust-scattered light, for very similar reasons. First, in both
instances the scattering geometry involves mainly large-angle
scattering, which does not lead to high scattered-light
intensities. Second, the illuminating radiation field in both cases
contains both a strong far-UV component and a strong optical component
(see Witt \& Johnson (1973) for a spectrum of the interstellar radiation field), a condition which appears to be associated with
highly efficient production of ERE. The interstellar medium giving
rise to the high-$|b|$ Galactic cirrus is optically thin in the far-UV,
where the ERE is initiated, as well as at the ERE wavelength range
itself. This, coupled with  a reasonably accurate knowledge of the
spectrum of the exciting interstellar radiation field
\citep{Mathis83}, allows a reliable estimate of the ERE intensity per
hydrogen atom along an average line of sight to be made, as well as an
estimate of a lower limit of the photon conversion efficiency of the
ERE process \citep{Gordon98}.

\citet{Gordon98} found an ERE intensity of $(1.43 \pm 0.31)
\times 10^{-29}$~ergs~s$^{-1}$ $\AA^{-1}$~sr$^{-1}$~H-atom$^{-1}$ for the
Galactic cirrus at $|b| > 20$~degrees, independently confirmed by
\citet{Szomoru98}.  With the well-established observed relation
between hydrogen column density and extinction in the Galaxy
\citep{Diplas94}, \citet{Gordon98} estimated that ($10 \pm 3$)\% of all
photons that are absorbed from the interstellar radiation field by
interstellar dust must be absorbed by the ERE carrier to produce the
observed ERE intensity with a conversion efficiency of 100\% on a
photon-per-photon basis. If, as our current results suggest, far-UV
photons are required to initiate the ERE, we must first consider the
far-UV portion of the interstellar radiation field. If we consume
photons through absorption by the ERE carrier, starting at the 91.2~nm
limit set by the interstellar hydrogen opacity and proceed toward
longer wavelengths, we find that we must use all photons shortward of
130~nm in wavelength or with energies down to 9.5~eV. This limit is
indicated by a vertical dashed line in Figure~\ref{fig_ere_limits},
labeled 100\%. We note that this limit violates the ERE initiation
limit of 10.5~eV set by the opacity constraint found in NGC~7023. If
we assume a more realistic conversion efficiency of 50\% for the
photons absorbed by the ERE carriers, all photons absorbed between
91.2~nm and 185~nm in wavelength would be required to generate the
observed ERE intensity. This second limit is shown in
Figure~\ref{fig_ere_limits} as a dashed vertical line labeled 50\%.

Therefore, in close analogy to the case of the Red Rectangle discussed
earlier, we find that that the interstellar radiation field does not
contain enough photons with energies $E > 10.5$~eV to pump any ERE
carriers sufficiently to generate the ERE intensities observed from
the high-$|b|$ Galactic cirrus. Again, the problem would meet with a
simple solution, if the high-energy photons are used solely to produce
and maintain a population of ERE carriers which can then be pumped by
the far more abundant optical/near-UV photons of the interstellar
radiation field. Thus, a two-step process of ERE excitation is also
being suggested by the data available for the high-$|b|$ diffuse
interstellar medium.

\section{Conclusions}

We summarize our conclusions:

\begin{enumerate}

\item We have imaged sharp molecular cloud edges in the NW PDR of the
reflection
nebula NGC~7023 in the light of the ERE band and in the light of
emission from the H$_2$~1-0~S(1) transition, using the HST ACS and
HST NICMOS instruments, respectively.  The cloud edges appear as
narrow filaments with widths as small as 0.3 arcsec in both bands,
while the appearance of the same cloud edges is broader and more
diffuse in images showing dust-scattered radiation at H$\alpha$ and in
the z-band.

\item In general, every H$_2$ filament is matched by a corresponding
ERE filament, while the reverse correlation does not hold to the same
degree. We interpret this one-sided correlation as resulting from a
density-dependent conversion efficiency of far-UV photons into near-IR
1-0~S(1) photons. In lower-density regions, absorption of far-UV
photons by dust dominates over the absorption by H$_2$ molecules,
leading to ERE filaments without corresponding H$_2$ filaments, while
in higher-density regions absorptions of far-UV photons by dust and by
H$_2$ molecules are comparable.

\item By comparing the widths of H$_2$ filaments with the
corresponding widths of the cloud edges in the z-band, we estimated
the wavelength-dependence of the dust extinction in the molecular
clouds comprising the NW PDR of NGC~7023. Assuming that the extinction
follows the R$_V$-dependent relationship developed by CCM, we found a
value of R$_V \approx 5.6$.

\item By comparing the widths of the ERE filaments with the
corresponding widths of the cloud edges in the z-band, we found the
ERE filaments on average about 10\% wider than the corresponding
H$_2$ filaments.  This indicates that the dust opacity encountered by
photons that initiate the ERE is only slightly smaller than the dust
opacity encountered by photons exciting the H$_2$ molecules. Given
that the H$_2$ emission in NGC~7023 is excited by far-UV photons with
$\lambda > 110.8$~nm, the CCM-extinction curve of R$_V \approx 5.6$
identifies two regions of the spectrum where the dust opacity
matches the required value, the range around the mid-UV extinction
bump (250~nm to 189~nm) and the region shortward of 118~nm.

\item We used the observation by \citet{Darbon99} that stars with
T$_\mathrm{eff} \leq 7000$~K do not excite ERE to eliminate the possibility
that ERE is initiated by mid-UV photons in the 250~nm to 186~nm
range. This leaves far-UV photons at wavelengths shortward of 118~nm
($\sim$10.5~eV) as the ultimate source of ERE excitation.

\item We provided arguments showing that all existing ERE models fail in the face of this new requirement.

\item With the examples of two environments where the ERE excitation
is highly efficient and where the number of ERE photons has been
determined, we demonstrated that the number of far-UV photons with
energies in excess of 10.5~eV is insufficient to generate the number
of observed ERE photons, assuming reasonable values for the photon
conversion efficiencies.

\item We concluded that the ERE excitation must therefore be a
two-step process. The first step requires $E > 10.5$~eV photons to
produce the actual ERE carrier, probably by a process of
photo-ionization or photo-dissociation of a suitable precursor. A
modest far-UV flux would then suffice to maintain a population of ERE
carriers thus created. The second step of the ERE excitation then
requires efficient pumping of the ERE carrier by abundant
optical/near-UV photons with energies above the ERE band-gap of
$\approx 2$~eV.

\item A possible ERE carrier must therefore be a system that (1) has
an ionization or dissociation potential in excess of ~10.5~eV, that (2) exhibits
strong absorption in the optical/near-UV spectral region, and that (3)
is capable of efficient photo-luminescence in the wavelength range of
the ERE band.

\item We suggest that PAH di-cations with masses $\leq \sim500$~amu appear
to meet the first two of these three requirements. Laboratory
experiments designed to study the possibility of fluorescence by PAH
di-cations are needed to test their potential as ERE sources.
\end{enumerate}

\acknowledgements
Support for program \#9471 was provided by NASA through a grant from
the Space Telescope Science Institute, which is operated by the
Association of Universities for Research in Astronomy, Inc., under
NASA contract NAS 5-26555. Additional financial support for this work
was provided through NSF grant 0307307 to the University of
Toledo. RHX thanks HPCVL at Queen's University for the use of its
parallel supercomputing facilities.

\end{document}